\documentclass[12pt,draftclsnofoot,onecolumn]{IEEEtran}
\usepackage{amsfonts}
\usepackage{cite,graphicx,amsmath,amsthm}
\usepackage{subfigure}
\usepackage{fancyhdr}
\usepackage{dsfont}
\usepackage{array,color}
\usepackage{bm}
\usepackage{float}
\usepackage{algorithm}
\usepackage{algpseudocode}
\usepackage{multirow}
\usepackage{booktabs}
\usepackage{multirow}
\usepackage{makecell}
\usepackage{graphicx}
\usepackage{float}
\usepackage{subfigure}
\usepackage{ragged2e}
\usepackage{booktabs}
\usepackage{bbding}
\usepackage{multirow}
\usepackage{makecell}
\usepackage{amsmath,amssymb,amsfonts}
\usepackage[]{caption2}
\usepackage[table,xcdraw]{xcolor}
\usepackage{cite}
\allowdisplaybreaks[4]

\newtheorem{lemma}{Lemma}

\newtheorem{proposition}{Proposition}

\newtheorem{corollary}{Corollary}

\newtheorem{property}{Property}

\newtheorem{remark}{Remark}

\newtheorem{claim}{Claim}

\def\BibTeX{{\rm B\kern-.05em{\sc i\kern-.025em b}\kern-.08em
    T\kern-.1667em\lower.7ex\hbox{E}\kern-.125emX}}

\begin{document}
\title{{
Wireless Sensing With Deep Spectrogram Network and Primitive Based Autoregressive\\Hybrid Channel Model}}
\author{
Guoliang Li$^{*}$, Shuai Wang$^{*}$, Jie Li$^{*}$, Rui Wang$^{*}$,\\Xiaohui Peng$^{\diamond}$, and Tony Xiao Han$^{\diamond}$\\
        $^{*}$Southern University of Science and Technology, Shenzhen 518055, China\\
        $^{\diamond}$Huawei technologies, Co. Ltd., Shenzhen, China\\
E-mail: ligl2020@mail.sustech.edu.cn, wangs3@sustech.edu.cn, lij2019@mail.sustech.edu.cn, wang.r@sustech.edu.cn, pengxiaohui5@huawei.com, tony.hanxiao@huawei.com
        }

\maketitle
\vspace{-0.5in}
\begin{abstract}
Human motion recognition (HMR) based on wireless sensing is a low-cost technique for scene understanding.
Current HMR systems adopt support vector machines (SVMs) and convolutional neural networks (CNNs) to classify radar signals.
However, whether a deeper learning model could improve the system performance is currently not known.
On the other hand, training a machine learning model requires a large dataset, but data gathering from experiment is cost-expensive and time-consuming.
Although wireless channel models can be adopted for dataset generation, current channel models are mostly designed for communication rather than sensing.
To address the above problems, this paper proposes a deep spectrogram network (DSN) by leveraging the residual mapping technique to enhance the HMR performance.
Furthermore, a primitive based autoregressive hybrid (PBAH) channel model is developed, which facilitates efficient training and testing dataset generation for HMR in a virtual environment.
Experimental results demonstrate that the proposed PBAH channel model matches the actual experimental data very well and the proposed DSN achieves significantly smaller recognition error than that of CNN.
\end{abstract}


\IEEEpeerreviewmaketitle
\section{Introduction}

Wireless sensing is a promising technique for addressing the safety issues in intelligent transportation systems \cite{eldar,zijian}.
For instance, the technique is able to detect running children in the underground parking garage.
Wireless sensing can be classified into model-driven sensing \cite{localization} and data-driven sensing \cite{vcchen,linghao,linghao2}.
Localization is one of the typical applications of the former, where the locations of mobile devices are computed from a few measurements (e.g., difference of time of arrivals) of the radio frequency (RF) signals via geometry relationship \cite{localization}.
However, the data-driven sensing is usually much more complicated. For example, in human motion recognition (HMR), motions cannot be determined directly via the measurements of RF signals. Instead, the machine learning techniques are adopted to extract useful information 
by comparing with historical training samples \cite{vcchen}.

Specifically, the micro-Doppler features of the non-rigid body movements are first extracted from the received RF signals in \cite{vcchen}.
Then, these features are represented by the time-frequency spectrograms (i.e., images) and machine learning approaches are applied for image classification.
For example, in \cite{linghao}, the short time Fourier transform (STFT) is adopted to generate spectrograms and a support vector machine (SVM) is applied to the spectrograms, which achieves a classification error smaller than $10\%$.
By replacing the SVM with a convolutional neural network (CNN), it is shown in \cite{linghao2} that a smaller classification error is achievable.
However, it is unclear whether the system performance could be further improved by leveraging deeper neural networks such as residual network (ResNet) \cite{ResNet}.

On the other hand, training a machine learning model requires a large dataset.
However, data gathering from experiment is environmental and task dependent and thus extremely time consuming. Moreover, manual annotations are also necessary.
To reduce the costs of dataset collection, one promising approach is to generate various scenarios and massive labeled data via a wireless sensing channel model.
Nevertheless, wireless sensing is fundamentally different from wireless communications and the existing channel models may not be suitable for the tasks of wireless sensing.
In particular, based on our experimental results, it is found that a wireless sensing channel model should be consistent in spatial, time and micro-Doppler domains, while maintaining sensing uncertainty.
Unfortunately, none of the existing channel models \cite{ar,channelreview,primitive,ray,METIS,3gpp,IEEE,quadriga,schannel} satisfy the above features.

In summary, there are two main challenges in this paper: 1) Lack of a deep human motion recognition model, and 2) Lack of a wireless sensing channel model.
To address the above challenges, this paper proposes a deep spectrogram network (DSN) for high-quality human motion recognition and a primitive based autoregressive hybrid (PBAH) channel model to facilitate efficient training and testing dataset generation for HMR in a virtual environment. 
The contributions of this paper are summarized below:

\begin{itemize}
    \item[1)] \textbf{Development of a deep spectrogram network}.
    The DSN employs singular value decomposition (SVD) for data cleaning, STFT for data transformation, and deep residual learning for data classification, which achieves smaller classification error than that of CNN.

    \item[2)] \textbf{Development of a primitive based autoregressive hybrid channel model}.
    The PBAH channel model is a simulator that provides vivid sensing datasets and supports all the features required by HMR.
    The channel evolution rate of the PBAH model is fine-tuned via Kullback-Leibler (KL) divergence minimization.
    Experimental results show that the PBAH channel model matches the actual experimental data very well.

\end{itemize}

\section{Problem Statement}

A wireless sensing system placed in a conference room as shown at the left of Fig.~1 is considered.
The environment consists of a target person and a few static objects (e.g., walls).
The system aims to recognize human motions (e.g., standing, walking, running) via radar signal transmission and processing.
The radar is equipped with a single transmit antenna and a single receive antenna, but multi-antenna settings are equally valid.
Specifically, the radar transmits one frequency-modulated continuous wave (FMCW) every $T$ miliseconds.
Each signal lasts for $T_0$ (i.e., sweep time of FMCW) and the total number of transmitted FMCWs is set to $C$.
Hence the total sensing time is $TC$.
Let $s_i(t)$ denote the $i$-th FMCW ($1\leq i \leq C$).
The received signal at the radar is
\begin{align}
r_i(t,m)=h_i(t,m)\ast s_i(t)+n_i(t),  \label{rt}
\end{align}
where $h_i(t,m)$ is the channel impulse response from radar transmitter to receiver if the target person takes the $m$-th motion ($m\in\mathcal{M}$ with $\mathcal{M}=\{1\cdots,M\}$, $M$ is the number of human motions) and $n_i(t)$ is the additive white Gaussian noise (AWGN) at the radar receiver.
After collecting all the signals, the radar forward $\{r_i(t,m)\}_{i=1}^C$ to a processor (shown at the right of Fig.~1) for human motion recognition.
For the above system, two fundamental problems are: 1) How to design the machine learning model for processing the signals $\{r_i(t,m)\}_{i=1}^C$? 
2) How to generate a virtual dataset $\{\{r_i(t,m)\}_{i=1}^C,m\}\}_{m\in\mathcal{M}}$ for training learning models such that trained model has the consistent performance with the one trained from experiment data? 
Since $s_i(t)$ in \eqref{rt} is fixed and the power of $n_i(t)$ in \eqref{rt} is small, the key to generate $r_i(t,m)$ is to model the wireless channels $h_i(t,m)$.
In the following two sections, the proposed DSN model and the PBAH channel model will be elaborated to address the problems.

\section{Proposed Deep Spectrogram Network}

\begin{figure}[!t]
\centering
		\includegraphics[width=0.8\textwidth]{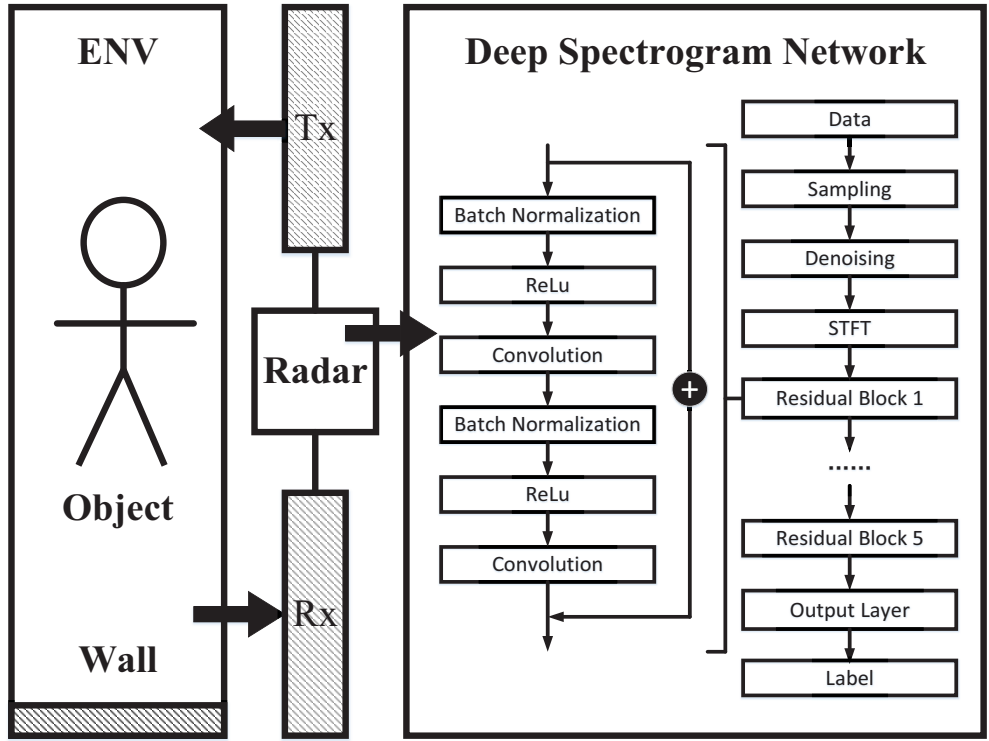}
	\caption{Illustration of the system model and the architecture of DSN.}
	\label{simulation} 
\end{figure}

The input of DSN (shown in Fig.~1) is the signals $\{r_i(t,m)\}_{i=1}^C$ and the output is the human motion category $\widehat{m}$.
The DSN consists of the following data sampling, cleaning, transformation, and classification steps.

\textbf{Data Sampling}.
The received signal $r_i(t,m)$ in \eqref{rt} is discretized into a vector $\mathbf{x}_i(m)\in\mathbb{C}^{L}$.
Since $r_i(t,m)$ lasts for $T_0$, the length of $\mathbf{x}_i(m)$ is $L=T_0f_s$, where $f_s$ is the sampling rate in Hz.
For example, if $T_0=1$ us and $f_s=100$ MHz, then we have $L=100$.
Since the radar transmits $C$ waves, the vectors $\mathbf{x}_1(m),\cdots,\mathbf{x}_C(m)$ are concatenated into a matrix $\mathbf{X}(m)\in\mathbb{C}^{L\times C}$.

\textbf{Data Cleaning}.
The matrix $\mathbf{X}(m)$ is a superposed signal consisting of the useful signals reflected from human and interference signals reflected from walls (or other objects).
To extract the useful information, we compute $\widetilde{\mathbf{X}}(m)=\mathrm{conj}\left(\mathrm{dechirp}\left(\mathbf{X}(m), \widetilde{\mathbf{s}}\right)\right)$ and $\widetilde{\mathbf{s}}$ is the discretization of $\{s_i(t)\}$ with sampling frequency $f_s$.
Then singular value decomposition is applied to $\widetilde{\mathbf{X}}(m)$, yielding $\widetilde{\mathbf{X}}(m)=\sum_{j=1}^Ra_j\mathbf{b}_j\mathbf{c}_j^H$, where $R$ is the rank of $\widetilde{\mathbf{X}}(m)$, $a_j$ is the power of the $j$-th basis, $\mathbf{b}_j,\mathbf{c}_j$ are the $j$-th basis vectors.
Removing the first $r-1$ components where $r$ is the pre-determined threshold, the denoised signal is $\mathbf{Y}(m)=\sum_{j=r}^Ra_j\mathbf{b}_j\mathbf{c}_j^H$.

\textbf{Data Transformation}.
Short-time Fourier transform (STFT) \cite{vcchen} with a sliding window function $w(i)$ of length $W$ is adopted to generate both time features and frequency features of $\mathbf{Y}(m)$.
The parameter $W$ controls the trade-off between time and frequency resolutions, and we use a Kaiser window with $0.128$ s ($W=128$ with a time step of $1~\mathrm{ms}$).
The time-frequency signal at time $m$ and the frequency $f$ is $\mathbf{Z}(m) = \mathrm{STFT}\left(\mathbf{y}(m)\right)$, where $\mathbf{y}(m)=\mathbf{1}^T\mathbf{Y}(m)$.
Moreover, the $\mathrm{STFT}$ function can be expressed as
$z(l, f) = \sum_{i=-\infty}^{+\infty} y(i)w(i-lW)\exp(-\mathrm{j}f i)$, 
where $z(l, f)$ is the element at the $l$-th column and $f$-th row of $\mathbf{Z}(m)$.

\textbf{Data Classification}.
The ResNet \cite{ResNet} is adopted as the backbone for the feature extraction, which consists of five identical residual blocks and a softmax classification layer.
Each residual block consists of six layers: a batch normalization layer, a ReLu activation layer, a convolution layer, a batch normalization layer, a ReLu activation layer, and a convolution layer.
For each residual block, the difference between its input and output is computed and learnt.
The output label is the human motion category $\widehat{m}$.

\section{Proposed Benchmark Metrics}

Training a machine learning model requires a large dataset, but data gathering from experiment is cost-expensive and time-consuming.
Wireless channel models can be exploited for dataset generation.
However, for simulation based dataset generation method, one fundamental question is: What makes a ``good'' simulator?
Existing benchmark metrics such as power fading, time delay, and angle of arrival (departure) \cite[Table~IV]{channelreview} are designed for communication rather than sensing.
To exploit new metrics for wireless sensing, a USRP RIO experiment is carried out in a conference room.
The experimental setup is shown in Fig.~2 and the experimental results are shown in Fig.~3.
It can be seen from Fig.~3 that a proper wireless sensing channel model should satisfy: \emph{micro-Doppler consistency} to represent non-rigid body movement (the periodic curve in the central pattern), and \emph{sensing uncertainty} to represent unpredictable object deformation (the blurs around the central pattern).
Besides, according to \cite{METIS}, a wireless channel model should also satisfy: \emph{spatial consistency} to represent geometry information, and \emph{time consistency} to represent the dynamics. Therefore, the following benchmark metrics for sensing are proposed:

\begin{figure}[!t]
	\centering
\includegraphics[width=0.95\textwidth]{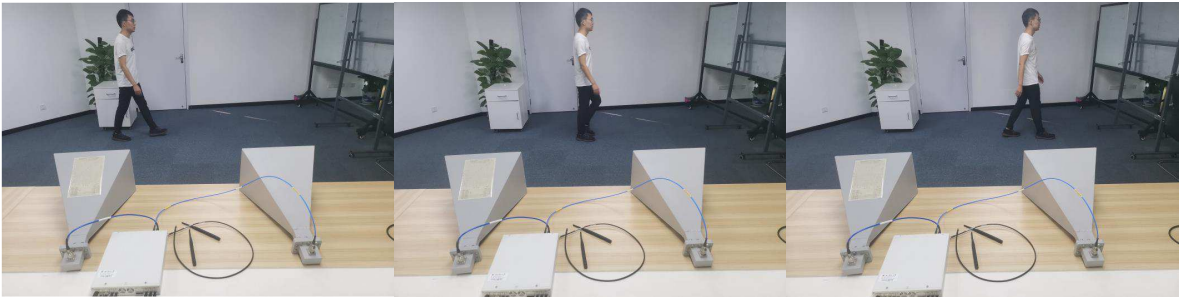}
		\caption{Sensing a walking human in a conference room using USRP RIO and two horn antennas. The experimental setting is as follows: the human is $1.75\,\mathrm{m}$ height; the walking speed is $1\,\mathrm{m/s}$; the carrier frequency is $3.5\,\mathrm{GHz}$; the sampling rate is $100\,\mathrm{MHz}$; the bandwidth is $50\,\mathrm{MHz}$; the sweep time of FMCW is $1\,\mathrm{\mu s}$.}
		\label{experiment} 
		\end{figure}

\begin{figure}[!t]
	\centering
 \subfigure{\includegraphics[width=0.32\textwidth]{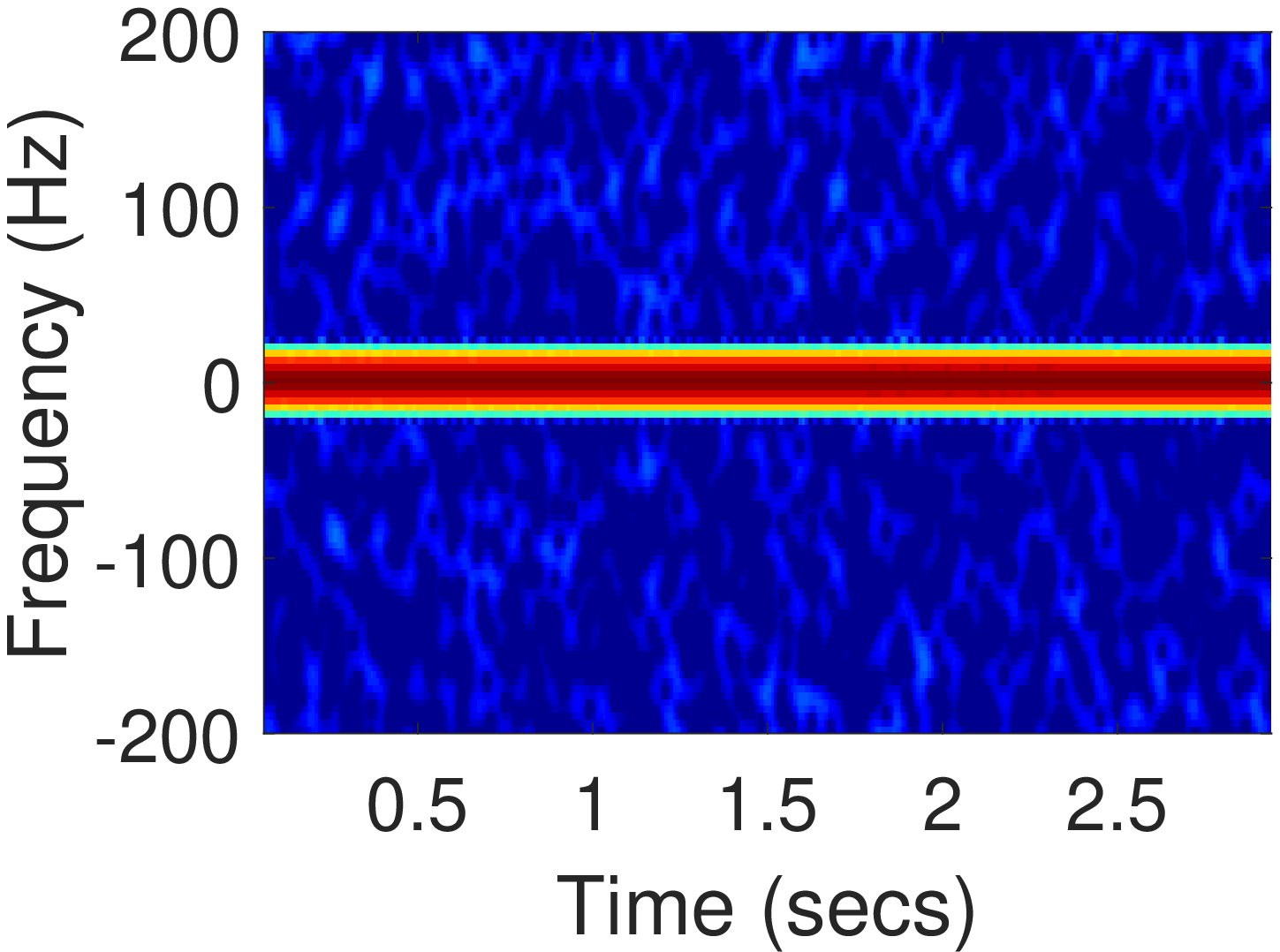}}
 \subfigure{\includegraphics[width=0.32\textwidth]{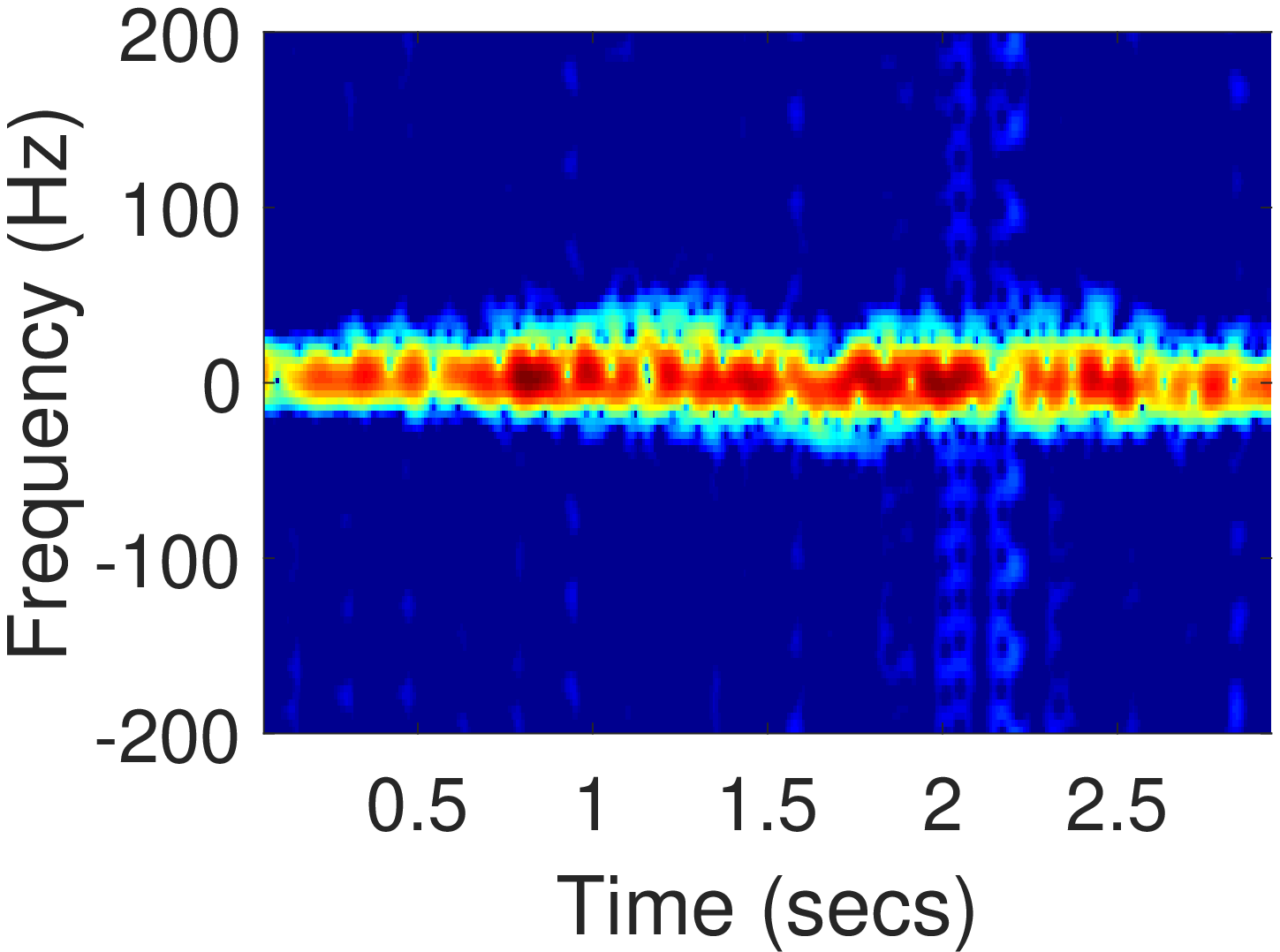}}
 \subfigure{\includegraphics[width=0.32\textwidth]{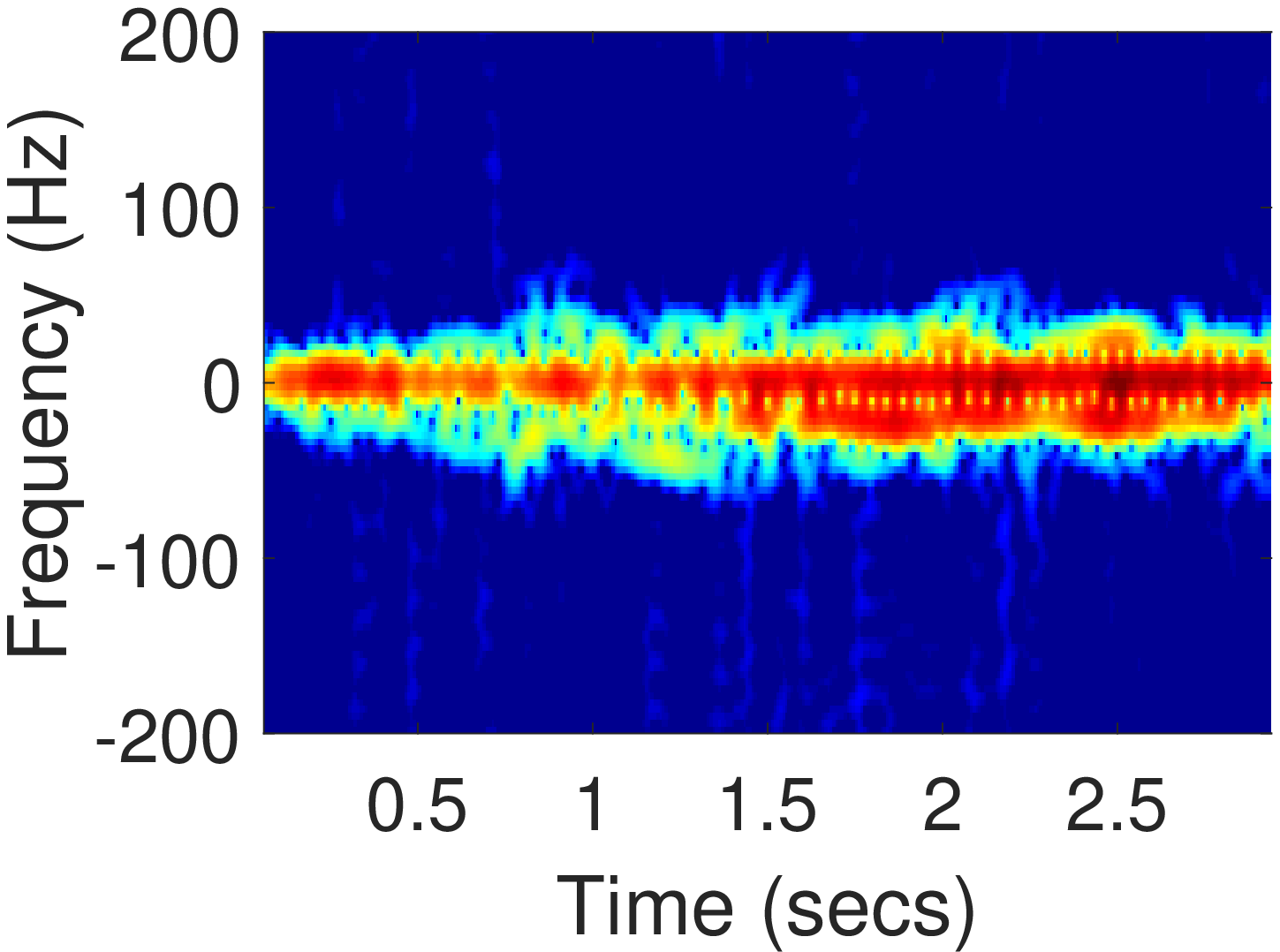}}
 \caption{The experimental results of wireless sensing of human motions. a) standing; b) walking towards wall; c) walking towards radar.
   }
	 \label{fig:subfig}
\end{figure}

\textbf{Spatial Consistency}. Channels are similar in closely located links.
    Large-scale spatial consistency refers to consistent power fading, delay spreads and angle spreads at two close locations.
    Small-scale spatial consistency refers to consistent delays and angles of rays at two close locations.
    Micro-scale spatial consistency refers to consistent phases at two close locations.
    Most existing channels support large-scale and small-scale spatial consistency \cite{METIS}, but do not support micro-scale spatial consistency.

\textbf{Time Consistency}. Channels are similar in close time points.
    Deterministic time consistency means that the channel evolves smoothly without discontinuities when the TX or RX or target moves (the middle curve in Fig.~3 is smooth) \cite{METIS}.
    Stochastic time consistency means that there is no absolutely static environment and the channel is changing due to unpredictable object deformation (e.g., crowds, swaying leaves, water flows).

\textbf{Micro-Doppler Consistency}. The micro-Doppler shift is consistent with the non-rigid body motions of humans.
    The fingerprints of motions (the periodic curve in Fig.~3) in the time-frequency spectrograms \cite{vcchen} can be used to distinguish various human activities such as walking and running.

\textbf{Sensing Uncertainty}. The random micro-Doppler shifts (the light points in Fig.~3) in the time-frequency spectrograms.
    Sensing uncertainty is caused by stochastic time evolution of wireless channels \cite{ar} and is a necessary feature to verify the robustness of sensing algorithms.

\textbf{Comparison with Existing Channel Models}.
As summarized in Table.~I, current channel models can be categorized into statistical model, deterministic model, and quasi-deterministic model.
However, none of the existing models can support all the features defined above.
Specifically, statistical models \cite{schannel} are intended for communication purpose but hardly supports any features for sensing applications.
Deterministic models \cite{ray,primitive} involve high computational complexities and do not support sensing uncertainty as the channels are fully determined by geometry.
Lastly, most standards in industry adopt quasi-deterministic channel models \cite{IEEE,3gpp,METIS,quadriga}.
Nonetheless, all those models are presented for communication purpose and not applicable for sensing.
In \cite{IEEE} and \cite{3gpp}, humans are modeled as random blockage and only affects power fading.
In \cite{METIS} and \cite{quadriga}, they describe the drifting method to generate the dynamic channel when the TX or RX moves.
But how to extend the drifting method to the case of non-rigid movement is not mentioned.
Moreover, all the models in \cite{IEEE,3gpp,METIS,quadriga} do not support modeling of sensing uncertainty.

\begin{table*}[!t]
\scriptsize
\caption{A Comparison of Existing and Proposed Channel Models}
\vspace{0.1in}
{

\centering
\begin{tabular}{|c|c|c|c|c|c|c|c|c|}
\hline
\hline
\textbf{Type} & \textbf{Literature} & \textbf{Methodology} & \textbf{Purpose} & \textbf{\begin{tabular}[c]{@{}c@{}}Compl. \end{tabular}}
& \textbf{\begin{tabular}[c]{@{}c@{}}Spatial \\ Consistency \end{tabular}}
& \textbf{\begin{tabular}[c]{@{}c@{}}Time \\ Consistency \end{tabular}}
 &
\textbf{\begin{tabular}[c]{@{}c@{}}Micro \\ Doppler \end{tabular}}
& \textbf{\begin{tabular}[c]{@{}c@{}}Sensing \\ Uncertainty \end{tabular}}
\\ \hline
\textbf{\begin{tabular}[c]{@{}c@{}}S\end{tabular}}                     & \cite{schannel}                         & cluster random process                                                             & commun.                                                               & +                                                          & L1                                                                & \XSolidBrush                                                                      & \XSolidBrush                                                                    & \XSolidBrush
\\ \hline
\textbf{\begin{tabular}[c]{@{}c@{}}D\end{tabular}}  & \cite{ray}             & ray tracing                                                              & both                                                                 & +++                                                        & L3                                                       & \Checkmark*                                                                     & \Checkmark                                                                   & \XSolidBrush
\\ \cline{2-9}
                                                                                     & \cite{primitive}                                           & primitive based                                                            & sensing                                                                 & ++                                                         & L3                                                                & \Checkmark*                                                                     & \Checkmark                                                                   &     \XSolidBrush                                                                                                                      \\ \hline
\multirow{5}{*}{\textbf{\begin{tabular}[c]{@{}c@{}}QD \end{tabular}}}  &
\begin{tabular}[c]{@{}c@{}}3GPP TR 38.901 \cite{3gpp}
\end{tabular} & map based hybrid                                                           & commun.                                                              & ++                                                               & L2                                                                & \Checkmark*                                                                     & \XSolidBrush                                                                    & \XSolidBrush
\\ \cline{2-9}
& QuaDRiGa \cite{quadriga}       & GBSM                                                               & commun.                                                                  & ++                                                                & L2                                                                & \Checkmark*                                                                      & \XSolidBrush                                                                    &\XSolidBrush
\\ \cline{2-9}
                                                                                     & IEEE 802.11 ay \cite{IEEE}       & QD based                                                               & commun.                                                                 & ++                                                                & L2                                                             & \Checkmark*                                                                      & \XSolidBrush                                                                    & \XSolidBrush
\\ \cline{2-9}
                                                                                     & METIS \cite{METIS}              & map based hybrid                                                              & commun.                                                                 & ++       & L2                                                                & \Checkmark*                                                                     & \XSolidBrush                                                                      & \XSolidBrush
\\ \cline{2-9}
                                                                                     & \textbf{Ours}                                                & PBAH                                                               & both                                                                 & ++          & L3                                                                & \Checkmark                                                                      & \Checkmark                                                                    & \Checkmark                                         \\ \hline
\hline
\end{tabular}
}
\vspace{0.1in}
\label{Table.related_work}
\hspace{1cm}

The symbol ``S'' means statistical, ``D'' means deterministic, ``QD'' means quasi-deterministic.
The symbol ``+'' means low complexity, ``++'' means moderate complexity, ``+++'' means high complexity.
The symbol ``L1'' means large-scale spatial consistency, ``L2'' means large-scale and small-scale spatial consistency, ``L3'' means all spatial consistency.
The symbol ``\checkmark'' means functionality supported, ``\XSolidBrush" means functionality not supported, ``\checkmark*'' means functionality partially supported.
\end{table*}

\section{Proposed Primitive Based Autoregressive Hybrid Channel Model}

This section develops a wireless sensing channel model satisfying all the features mentioned in Section~IV.
The index $m$ in $h_i(t,m)$ is omitted in this section.
Firstly, based on the number of refections, $h_i(t)$ can be decomposed into $h_i(t)=h_i^{[0]}(t)+h_i^{[1]}(t)+h_i^{[2]}(t)$,
where $h_i^{[0]}(t)$ is the direct channel from Tx to Rx, $h_i^{[1]}(t)$ is the first-order refection channel from Tx to objects to Rx, $h_i^{[2]}(t)$ is the higher-order refection channel from Tx to objects then to other objects and to Rx.
Since $h_i^{[0]}(t)\ast s_i(t)$ has no information about the target objects, the radar adopts self-interference cancelation techniques to remove it from $r_i(t)$.
On the other hand, the channel $h_i^{[2]}(t)$ consists of multiple refections of electromagenetic waves, and its large-scale fading is much larger than $h_i^{[1]}(t)$.
Therefore, it is reasonable to assume $h_2(t)\approx 0$. 
Then, the dominant channel is $h_i^{[1]}(t)$.

In this paper, the following hybrid channel model is adopted:
\begin{align}
&h_i(t)=u_i(t)+v_i(t), \label{PBH}
\end{align}
where $u_i(t)$ is the useful channel from Tx to target object to Rx and $v_i(t)$ is the interference channel from Tx to wall or other objects to Rx.
Specifically, the useful channel $u_i(t)$ is modeled using the primitive-based method as \cite{primitive}
\begin{align}
u_i(t)&
=\frac{A}{\sqrt{4\pi}}
\sum_{b=1}^B
  \frac{\sqrt{G_{b}(t)}}{D_b^2(t)}
    \mathrm{exp}\left(-\mathrm{j}\frac{2\pi f_c}{c}2D_b(t)\right)
   \mathrm{exp}\left(\mathrm{j}\varphi_{b}\right)
  \delta\left(t-\frac{2D_b(t)}{c}\right), \label{uit}
\end{align}
where $A$ is a constant related to antenna gain $P_t$, $G_b(t)$ is the scattering loss relate to radar cross sections at time $t$, $D_b(t)$ is the distance in m from the $b$-th primitive to the radar at time $t$, $f_c$ is the carrier frequency in Hz, $c=3\times 10^8\,\rm{m/s}$ is the speed of light, $\varphi_{b}\sim\mathcal{U}[-\pi,\pi]$ is the initial phase of the $b$-th ray, $\delta\left(\cdot\right)$ is indicator function.
The values of $D_b(t)$ and $G_{b}(t)$ are generated from computer animation.

On the other hand, the interference channel $v_i(t)$ is modeled by the autoregressive method as
\begin{align}
&v_i(t)=
\left\{
\begin{aligned}
&\Upsilon(t)
,~~~~~~~~~~~~~~~~~~~~~~~~~\,\quad\mathrm{if}~0\leq t\leq T_0
\\ &
\rho v_i(t-T_0)+(1-\rho)\Upsilon(t)
,\quad\mathrm{if}~t > T_0
\end{aligned}
\right.
, \label{vit}
\end{align}
where $\Upsilon(t)$ is the quasi-deterministic channel of IEEE 802.11ay, $\rho$ is the \emph{channel evolution rate }(a larger (smaller) $\rho$ leads to a slower (faster) channel evolution rate), and $T_0$ is the coherent time (in s).
Specifically, according to \cite{IEEE}, $\Upsilon(t)$ is modeled as
\begin{align}
\Upsilon(t)&=
\sum_{n=1}^N\sqrt{H_n}\frac{\lambda}{4\pi (D_0+\tau^{\mathrm{cluster}}_{n}c)}
\times
\left[\sum_{m=1}^Ma_{n,m}\mathrm{exp}\left(\mathrm{j}\phi_{n,m}\right)
  \delta(t-\tau^{\mathrm{ray}}_{n,m})\right],
\end{align}
where $H_n$ is reflection loss, $\lambda$ is wave length (in m), $\tau^{\mathrm{cluster}}_{n}$ is the time delay (in s) of the $n$-th cluster (obtained from ray-tracing), $\tau^{\mathrm{ray}}_{n,m}$ is the time delay (in s) of $m$-th ray in the $n$-th cluster (obtained from Poisson distribution), $a_{n,m}$ is amplitude of the $m$-th ray in the $n$-th cluster (obtained from Rayleigh distribution), $\phi_{n,m}$ is initial phase of the $m$-th ray in the $n$-th cluster (obtained from uniform distribution).
Since the proposed hybrid channel model is based on the primitive-based method and the autoregressive method, the model is termed primitive-based autoregressive hybrid (PBAH) channel.

It can be seen from equation \eqref{vit} that $v_i(t)$ is a function of $\rho$.
Since $r_i(t)$ is a function of $h_i(t)$ as seen from \eqref{rt} and $h_i(t)$ is a function of $v_i(t)$ as seen from \eqref{PBH}, $\{r_i(t)\}_{i=1}^C$ is a function of $\rho$.
Therefore, the spectrogram of $\{r_i(t)\}_{i=1}^C$ is related to $\rho$ and let $\Xi(\mathbf{z}|\rho)$ denote the probability mass function (pmf) of the gray scale values $\mathbf{z}=[z_1,\cdots,z_E]^T)\in\mathbb{R}_+^E$ in spectrogram of $\{r_i(t)\}_{i=1}^C$ given $\rho$, where $E$ is the number of gray-scale intervals.
Moreover, let $\Psi(\mathbf{z})$ denote the pmf of the gray scale values in the spectrogram of the real data.
The $\rho$ can be chosen to minimize the following KL divergence:
\begin{align}
\mathop{\mathrm{min}}_{0\leq\rho\leq 1}\quad&\sum\limits_{j=1}^{E}\Psi(z_j)\left[\mathrm{log}\left(\frac{\Psi(z_j)}{\Xi(z_j|\rho)}\right)\right].
\label{fitting}
\end{align}
The above problem can be solved by one-dimensional brute-force search.
In practice, $\rho$ is different in different scenarios and the values of $\rho$ can be stored in a look-up table.

\begin{figure*}[!t]
	\centering
		 \subfigure[]{
	   \label{fig:subfig:b} 
	   \includegraphics[height=33mm]{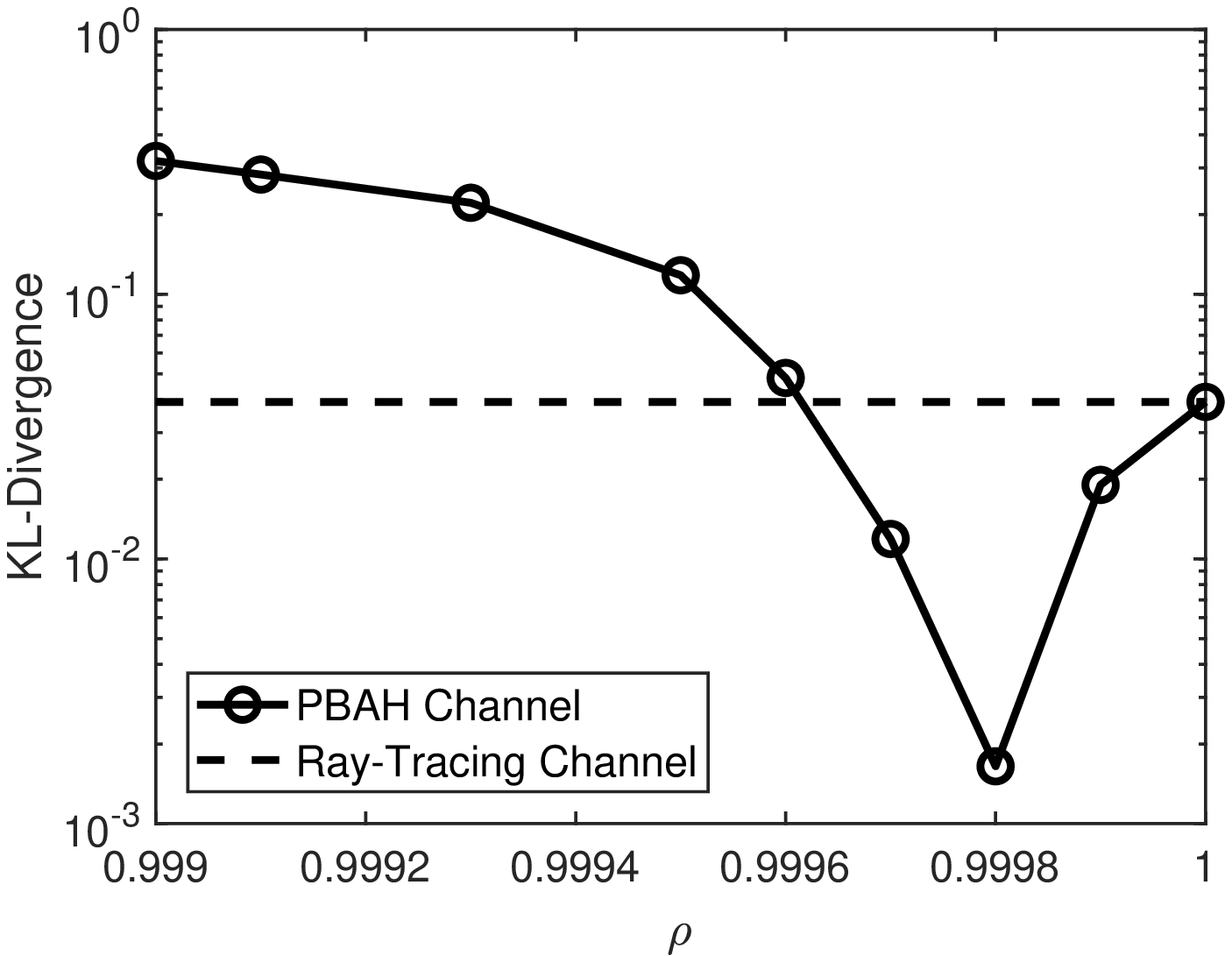}}
			 \subfigure[]{
	   \label{fig:subfig:c} 
	   \includegraphics[height=33mm]{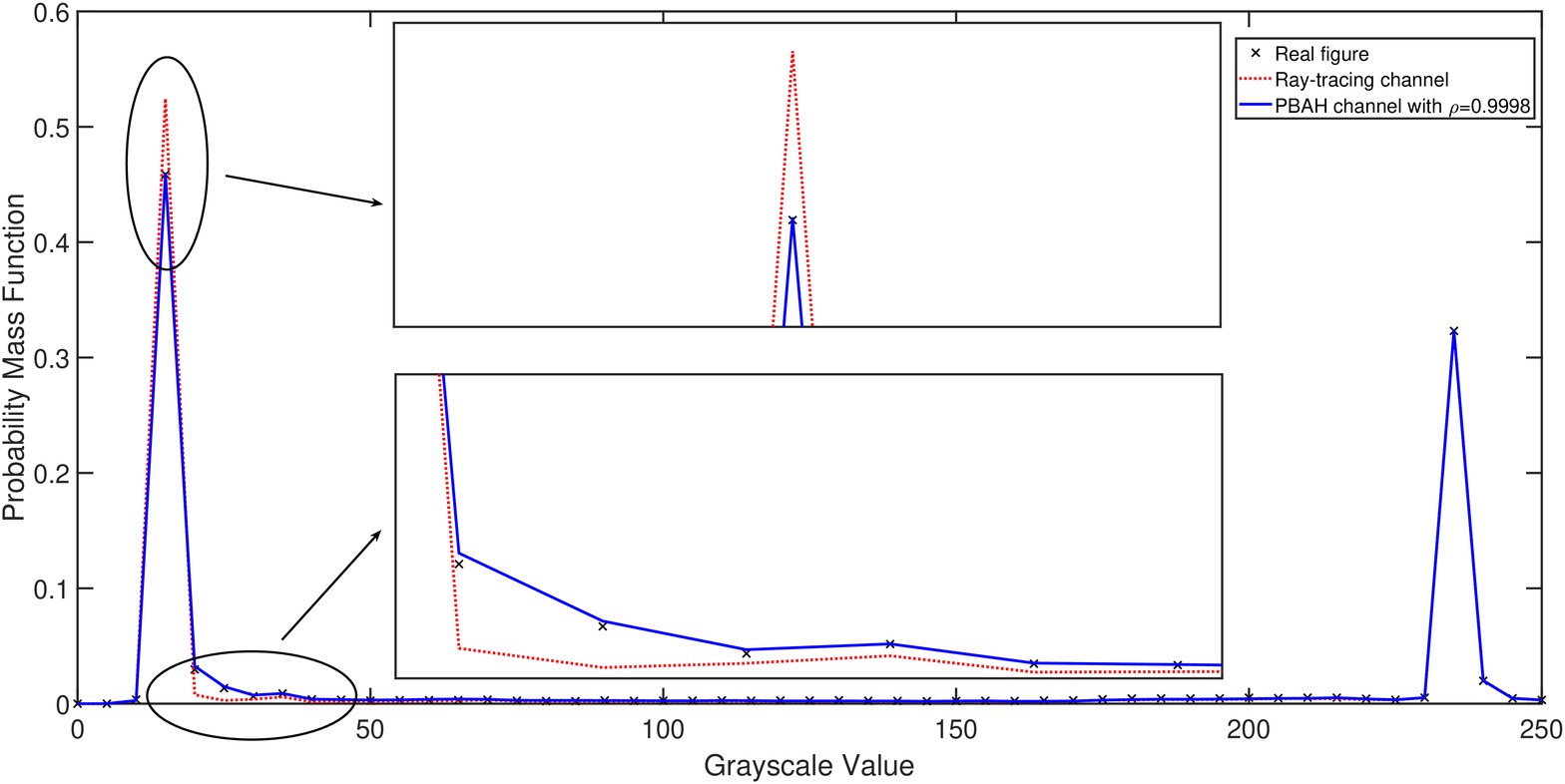}}	
		 \subfigure[]{
	   \label{fig:subfig:b} 
	   \includegraphics[height=33mm]{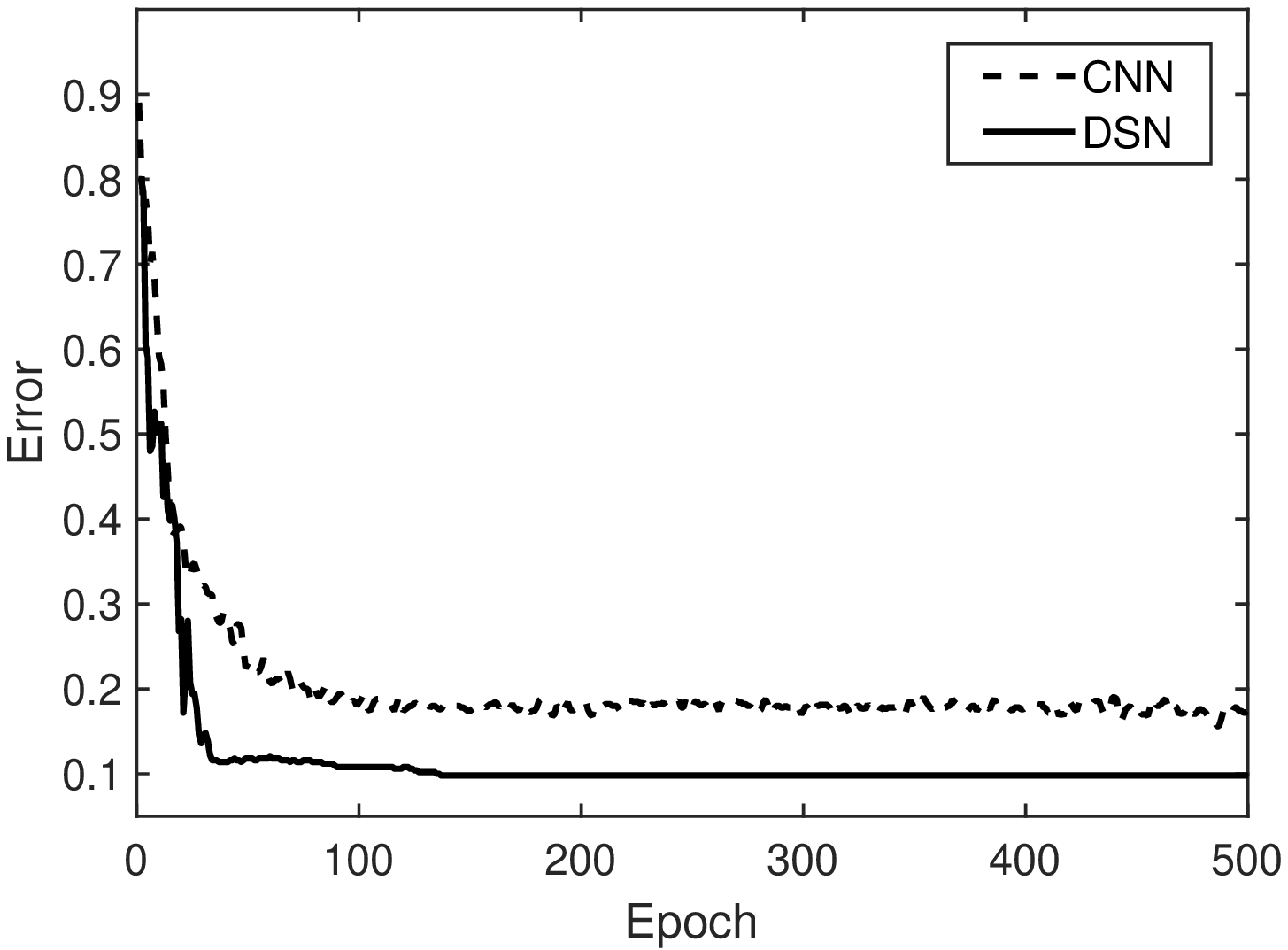}}
	
	 \caption{a) The KL divergence between the real data and simulated data under various $\rho$; b) Cmparison of pmf among the real data, the PBAH channel model, and the ray tracing channel model; c) Comparison between the DSN and the CNN.
   }
	 \label{fig:subfig} 
\end{figure*}

\section{Simulation Results and Discussions}

This section presents simulation and experimental results to verify the proposed channel and learning models.
We consider a conference room environment with size $(4.5\,\mathrm{m}$, $3\,\mathrm{m}$, $3\,\mathrm{m})$ (i.e., length, width, height).
Denoting the lower left corner of the room as $(0,0,0)$, the radar is located at $(1,1.5,1)$.
The carrier frequency is set to $f_c=3.5\,\mathrm{GHz}$ and the wavelength is $\lambda=0.0857\,\mathrm{m}$.
The transmit power at the radar is set to $P=1\,\mathrm{W}$ and the total bandwidth is $B_W=50\,\mathrm{MHz}$.
The transmission of FMCW is directional and the antenna gain is $P_t=25\,\mathrm{dB}$.
The time duration between consecutive FMCWs is $T=1~\rm{ms}$.
The sampling frequency and sweep time of FMCW are $f_s=100\,\mathrm{MHz}$ and $T_{0}=1\,\mathrm{\mu s}$, respectively.
The noise power is set to $-100\,\mathrm{dBm}$.
The phased array system toolbox of Matlab is adopted to generate the component $u_i(t)$ in $h_i(t)$ of \eqref{PBH}, and the IEEE 802.11ay channel simulator is adopted to generate the component $v_i(t)$ in $h_i(t)$ of \eqref{PBH}.

To verify the PBAH channel model, we consider the case of $C=3000$.
The object is assumed to be an adult, who walks along the y-axis from position $(4.2,0,0)$ to $(4.2,3,0)$ for $3\,\mathrm{s}$ with a speed of $1\,\mathrm{m/s}$.
The adult has $B=16$ primitives and the parameters $\{G_b(t),D_b(t)\}_{b=1}^B$ are computed from the positions of all primitives using the phased array system toolbox.
The number of FMCWs is set to $C=3000$.
Under the above settings, the KL divergence in \eqref{fitting} versus the value of $\rho$ is shown in Fig.~4a.
It can be seen from Fig.~4a that the KL divergence is very sensitive to the value of $\rho$ and the optimal solution $\rho^*=0.9998$.
With $\rho^*=0.9998$, the pmfs of the spectrograms for the real data, the PBAH channel model, and the ray tracing channel model are compared in Fig.~4b.
It can be seen that the ray-tracing channel model mismatches the real data when the gray-scale value is between $10$ and $50$.
This region corresponds to the blurs around the central pattern.
On the other hand, the proposed PBAH channel model matches the real data well and the KL divergence is one order of magnitude smaller than that of ray-tracing channel.
This corroborates the results of Fig.~5, where Fig.~5a and Fig.~5b look very similar (there exist a lot of burrs around the central pattern), while Fig.~5a and Fig.~5c look different (the middle pattern in Fig.~5c has very smooth edges).
Hence, the proposed PBAH channel is a better simulator of the real-word sensing channel than the ray-tracing channel.

To evaluate the performance of the proposed DSN, we consider the human motion dataset generated by PBAH channel model with the number of FMCWs $C=1000$.
The dataset contains five human motions: child/adult standing, child walking, child pacing, adult walking, adult pacing.
The adult is $1.75\,\mathrm{m}$ height and the child is $1\,\mathrm{m}$ height.
The speed of standing, walking, and pacing are $0\,\mathrm{m/s}$, $1\,\mathrm{m/s}$ and $0.5\,\mathrm{m/s}$, respectively.
The training of DSN is implemented via Momentum optimizer with a learning rate of $0.06$ and a mini-batch size of $500$.
After training for $500$ epochs, the training loss converges.
Then we test the trained model on a dataset with $500$ unseen samples, and compute the corresponding recognition accuracy.
It is observed from Fig.~4c that with the recognition error of the proposed DSN is significantly smaller that of CNN ($9.8\%$ versus $17.4\%$).

\begin{figure}[!t]
	\centering
		 \subfigure[]{
	   \label{fig:subfig:b} 
	   \includegraphics[width=0.31\textwidth]{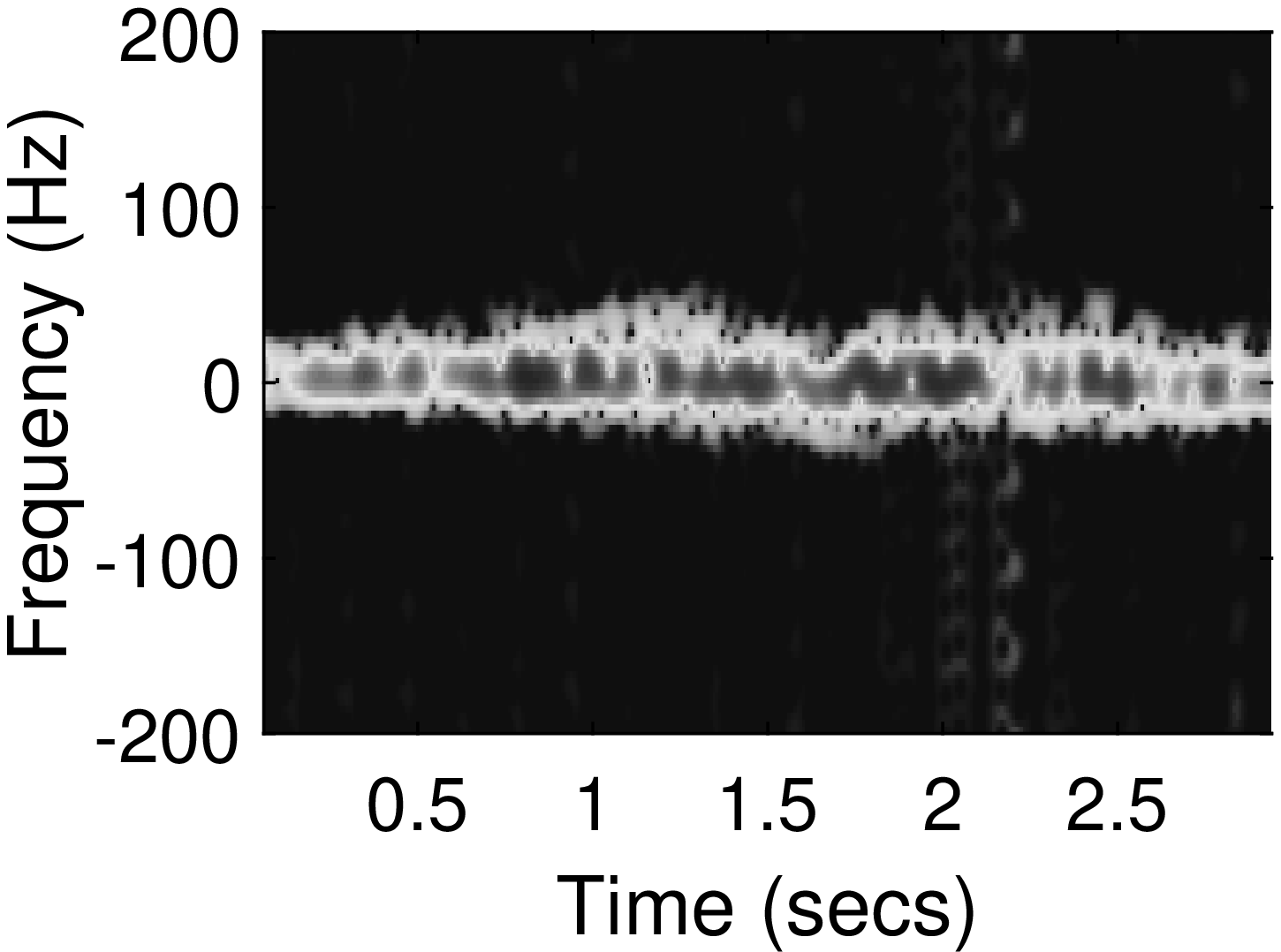}}
			 \subfigure[]{
	   \label{fig:subfig:c} 
	   \includegraphics[width=0.31\textwidth]{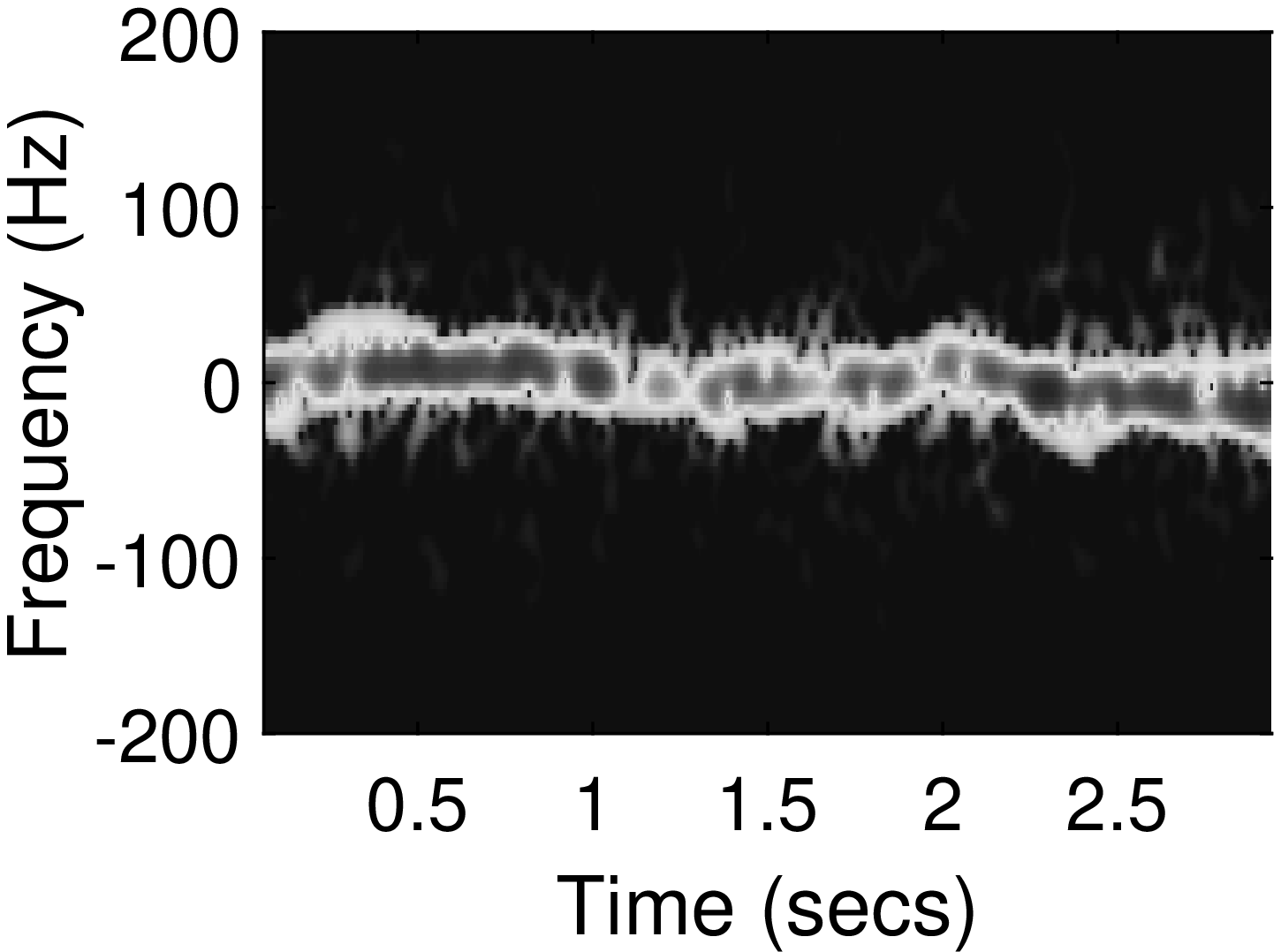}}
		 \subfigure[]{
		   \label{fig:subfig:c} 
	   \includegraphics[width=0.31\textwidth]{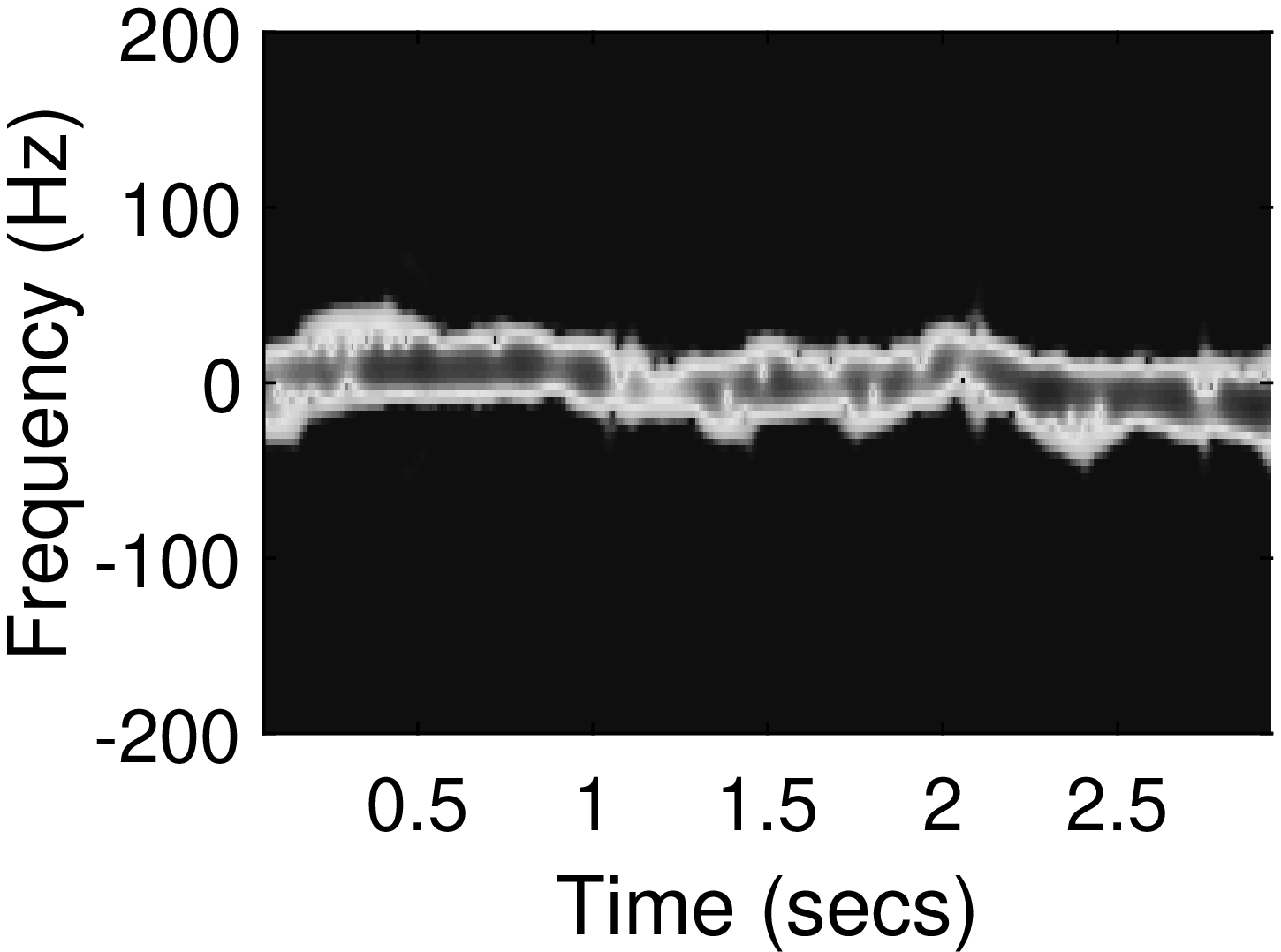}}
	 \caption{The comparison of spectrograms among the experimental data, the simulation dataset generated by PBAH with $\rho = 0.9998$, and the simulation dataset generated by ray-tracing. a) Actual data; b) PBAH channel model with $\rho = 0.9998$; c) Ray-tracing channel model.
   }
	 \label{fig:subfig} 
\end{figure}

\section{Conclusion}

This paper studied the deep learning and wireless channel models for human motion recognition.
A DSN was proposed to enhance the recognition performance.
To reduce the experimental costs, four benchmark metrics for wireless sensing were proposed and a PBAH channel model was developed satisfying all the metrics.
Finally, experimental results have shown that the proposed PBAH channel matches the actual data very well and the proposed DSN model achieves much higher recognition accuracy than that of existing networks.


\begin{thebibliography}{20}

\bibitem{eldar} D.~Ma, N.~Shlezinger, T.~Huang, Y.~Liu, and Y.~C.~Eldar, ``Joint radar-communication strategies for autonomous vehicles: Combining two key automotive technologies,'' \emph{IEEE Signal Process. Mag.}, vol.~37, no.~4, pp.~85--97, Jul.~2020,

\bibitem{zijian} Z. Zhang, S. Wang, Y. Hong, L. Zhou, and Q. Hao, ``Distributed dynamic map fusion via federated learning for intelligent networked vehicles,'' in \emph{Proc. IEEE ICRA}, Xi’an, China, May 2021.

\bibitem{localization} F.~Liu, C.~Masouros, A.~P.~Petropulu, H.~Griffiths, and L.~Hanzo, ``Joint radar and communication design: Applications, state-of-the-art, and the road ahead,'' \emph{IEEE Trans. Commun.}, vol. 68, no. 6, pp. 3834--3862, Jun. 2020.

\bibitem{vcchen} V.~C. Chen, F.~Li, S.-S.~Ho, H.~Wechsler, ``Micro-Doppler effect in radar: Phenomenon, model, and simulation study,'' \emph{IEEE Trans.  Aerosp. Electron. Syst.}, vol.~42, no.~1, pp.~2--21, Jan.~2006.

\bibitem{linghao} Y.~Kim, and H.~Ling, ``Human activity classification based on micro-Doppler signatures using a support vector machine,'' \emph{IEEE Trans. Geosci. Remote Sens.}, vol. 47, no. 5, pp. 1328--1337, May 2009.

\bibitem{linghao2} Y.~Kim and T.~Moon, ``Human detection and activity classification based on micro-Doppler signatures using deep convolutional neural networks,'' \emph{IEEE Geosci. Remote Sens. Lett.}, vol. 13, no. 1, pp. 8-12, Jan. 2016.

\bibitem{ResNet} K.~He, X.~Zhang, S.~Ren and J.~Sun, ``Deep residual learning for image recognition'' in \emph{Proc. IEEE Conf. Comput. Vis. Pattern Recognit.}, pp. 770-778, 2016.

\bibitem{channelreview} C.-X.~Wang, J.~Bian, J.~Sun, W.~Zhang, and M.~Zhang, ``A survey of 5G channel measurements and models,'' \emph{IEEE Commun. Sur. Tut.}, vol.~20, no.~4, Fourth Quarter 2018.

\bibitem{METIS} V.~Nurmela et al., METIS Channel Models, document ICT-317669/D1.4, METIS, New York, NY, USA, Jul. 2015.

\bibitem{ar} A.~\"O.~Kaya, L.~J.~Greenstein, and W.~Trappe, ``Characterizing indoor wireless channels via ray tracing combined with stochastic modeling,'' \emph{IEEE Trans. Wireless Commun.}, vol.~8, no.~8, pp.~4165--4175, Aug.~2009.

\bibitem{schannel} A.~A.~M.~Saleh and R.~Valenzuela, ``A statistical model for indoor multipath propagation,'' \emph{IEEE J. Sel. Areas Commun.}, vol.~5, no.~2, pp.~128--137, Feb.~1987.

\bibitem{ray} S.~Y. Seidel and T.~S. Rappaport, ``Site-specific propagation prediction for wireless in-building personal communication system design,'' \emph{IEEE Trans. Veh. Technol.}, vol.~43, no.~4, pp.~879--891, Nov.~1994.

\bibitem{primitive} G.~Duggal, S.~Vishwakarma, K.~V.~Mishra, and S.~S.~Ram, ``Doppler-resilient 802.11ad-based ultrashort range automotive joint radar-communications system,'' \emph{IEEE Trans.  Aerosp. Electron. Syst.}, vol.~56, no.~5, pp.~4035--4048, Oct.~2020.

\bibitem{3gpp} 3GPP, ``Study on channel model for frequencies from 0.5 to 100 GHz,'' 3GPP TR 38.901 V16.1.0, Dec. 2019.

\bibitem{quadriga} S.~Jaeckel, L.~Raschkowski, K.~B\"orner, and L.~Thiele, ``A 3-D multi-cell channel model with time evolution for enabling virtual field trials,'' \emph{IEEE Trans. Antennas Propag.}, vol. 62, no. 6, pp. 3242~256, Jun. 2014.

\bibitem{IEEE} A.~Maltsev et al., Channel Models for IEEE 802.11ay, document 802.11-15/1150r9, IEEE, New York, NY, USA, 2016.

\end{thebibliography}
\end{document}